\documentclass[aps,prb,twocolumn,showpacs,superscriptaddress,amsmath,amssymb,floatfix,10pt]{revtex4-1}

\usepackage{graphicx}
\usepackage{mathtools}
\usepackage{pbox}
\usepackage{adjustbox}
\usepackage{multirow}
\usepackage{braket}
\usepackage[table]{xcolor}
\usepackage{booktabs}
\usepackage{chemformula}
\usepackage[T1]{fontenc}
\usepackage[breaklinks=true,colorlinks,citecolor=blue,linkcolor=blue,urlcolor=blue]{hyperref}
\usepackage{adjustbox}
\usepackage[bottom]{footmisc}
\usepackage{tablefootnote}
\usepackage{tikz}

\usepackage{graphicx}
\usepackage{float}
\usepackage{placeins}

\renewcommand{\arraystretch}{1.08}

\newcommand{\RRef}[1]{Ref. \onlinecite{#1}}

\newcommand{\Eq}[1]{Eq.~(\ref{#1})}

\newcommand{\Fig}[1]{Fig. \ref{#1}}

\newcommand{\etal}{{\it et al.}}

\begin{document}

\title{Adaptive Expansions of the Optimized Effective Potential in Physically Motivated Response Spaces}

\author{Gabriel Chirchir}
\email{gabrielchirchir@gmail.com}
\affiliation{Institute of Physics, Faculty of Physics, Astronomy and Informatics, \\ Nicolaus Copernicus University in Toru\'n, \\ ul. Grudzi\k{a}dzka 5, 87-100 Toru\'n, Poland}
 \affiliation{Institute of Advanced Studies, Nicolaus Copernicus University in Toru\'{n}, ul. Wile\'{n}ska 4, 87-100 Toru\'{n}, Poland} 

 \author{Aditi Singh}
\affiliation{Institute of Physics, Faculty of Physics, Astronomy and Informatics, \\ Nicolaus Copernicus University in Toru\'n, \\ ul. Grudzi\k{a}dzka 5, 87-100 Toru\'n, Poland}
 \affiliation{Institute of Advanced Studies, Nicolaus Copernicus University in Toru\'{n}, ul. Wile\'{n}ska 4, 87-100 Toru\'{n}, Poland} 

 \author{Yan Lukashevich}
\affiliation{Institute of Physics, Faculty of Physics, Astronomy and Informatics, \\ Nicolaus Copernicus University in Toru\'n, \\ ul. Grudzi\k{a}dzka 5, 87-100 Toru\'n, Poland}
 \affiliation{Institute of Advanced Studies, Nicolaus Copernicus University in Toru\'{n}, ul. Wile\'{n}ska 4, 87-100 Toru\'{n}, Poland}

\author{Igor Sawicki}
\affiliation{Institute of Physics, Faculty of Physics, Astronomy and Informatics, \\ Nicolaus Copernicus University in Toru\'n, \\ ul. Grudzi\k{a}dzka 5, 87-100 Toru\'n, Poland}
 \affiliation{Institute of Advanced Studies, Nicolaus Copernicus University in Toru\'{n}, ul. Wile\'{n}ska 4, 87-100 Toru\'{n}, Poland} 

\author{Bogumiła Jezierska}
\affiliation{Institute of Physics, Faculty of Physics, Astronomy and Informatics, \\ Nicolaus Copernicus University in Toru\'n, \\ ul. Grudzi\k{a}dzka 5, 87-100 Toru\'n, Poland}
 \affiliation{Institute of Advanced Studies, Nicolaus Copernicus University in Toru\'{n}, ul. Wile\'{n}ska 4, 87-100 Toru\'{n}, Poland}

\author{Subrata Jana}
\affiliation{Institute of Physics, Faculty of Physics, Astronomy and Informatics, \\ Nicolaus Copernicus University in Toru\'n, \\ ul. Grudzi\k{a}dzka 5, 87-100 Toru\'n, Poland}
 \affiliation{Institute of Advanced Studies, Nicolaus Copernicus University in Toru\'{n}, ul. Wile\'{n}ska 4, 87-100 Toru\'{n}, Poland} 

\author{Szymon \'Smiga}
\email{szsmiga@fizyka.umk.pl}
\affiliation{Institute of Physics, Faculty of Physics, Astronomy and Informatics, \\ Nicolaus Copernicus University in Toru\'n, \\ ul. Grudzi\k{a}dzka 5, 87-100 Toru\'n, Poland}
 \affiliation{Institute of Advanced Studies, Nicolaus Copernicus University in Toru\'{n}, ul. Wile\'{n}ska 4, 87-100 Toru\'{n}, Poland} 

\date{\today}

\begin{abstract}
The optimized effective potential (OEP) method provides an exact
framework for incorporating orbital-dependent exchange within
Kohn-Sham (KS) density-functional theory (DFT). Its practical
implementation, however, requires the representation of the
exchange-response potential in a suitable auxiliary space, for which
conventional choices are not necessarily adapted to the physical
structure of the OEP response. Here, we introduce a general
response-space strategy in which physically motivated response
functions from existing model exchange potentials are repurposed as
adaptive auxiliary directions for the OEP equation. Specifically,
response functions associated with the Becke-Johnson (BJ),
R\"as\"anen-Pittalis-Proetto (RPP), Gritsenko-van Leeuwen-van
Lenthe-Baerends (GLLB), Krieger-Li-Iafrate (KLI), and localized
Hartree-Fock (LHF) constructions are incorporated into compact
auxiliary spaces, while their coefficients are determined directly
from the projected OEP equation rather than fixed by the assumptions
of the underlying model potentials. This establishes a systematic
connection between model exchange potentials and finite-basis OEP:
the former provide physically informed response directions, whereas
the latter determines their system-dependent amplitudes. The
resulting framework encompasses one- and multidimensional response
spaces as well as occupied-orbital and occupied-pair representations,
without changing the underlying OEP condition. We show that these
physically adapted spaces can represent the dominant spatial
structures of the OEP exchange response with a substantially reduced
number of degrees of freedom compared with conventional auxiliary
expansions. The proposed approach therefore provides a general route
to constructing compact, adaptive representations of exchange-only
OEP potentials and offers a systematic framework for developing
low-dimensional OEP approximations from physically motivated model
response functions.

\end{abstract}

\maketitle

\section{Introduction}





The optimized effective potential (OEP) method provides a rigorous framework for incorporating orbital-dependent energy functionals into the Kohn-Sham (KS) density-functional theory (DFT) \cite{SharpHorton1953,TalmanShadwick1976,KummelKronik2008}. Of particular importance is the exchange-only OEP (OEPx) method,
in which the exchange energy is evaluated exactly from the occupied
KS orbitals
\begin{equation}
E_x[{\phi_{i\sigma}}] = 
-\frac{1}{2}
\sum_{\sigma}
\sum_{i,j}^{\mathrm{occ}}
\iint
\frac{
\phi_{i\sigma}^{*}(\mathbf r)
\phi_{j\sigma}(\mathbf r)
\phi_{j\sigma}^{*}(\mathbf r')
\phi_{i\sigma}(\mathbf r')
}{
|\mathbf r-\mathbf r'|
}
d\mathbf r d\mathbf r'.
\end{equation}
This expression has the same algebraic form as the Hartree--Fock (HF) exact-exchange (EXX) energy, but it is evaluated with orbitals generated by a local multiplicative KS potential. Consequently, OEPx and HF generally yield different orbitals and, therefore, different exchange energies, despite using the same EXX energy expression. OEPx can thus be regarded as the exact treatment of exchange within the local KS framework, rather than as a local representation of a fixed HF solution.

Typically, imposing stationarity of the total energy with respect to
variations in the local potential leads to a Fredholm integral
equation of the first kind,
\begin{equation}
\int
\chi_{s,\sigma}(\mathbf r,\mathbf r')
v_{x,\sigma}^{\mathrm{OEP}}(\mathbf r')
\,d\mathbf r'
=
\Lambda_{x,\sigma}(\mathbf r),
\end{equation}
where $\chi_{s,\sigma}$ is the static KS response function and
$\Lambda_{x,\sigma}$ contains occupied--virtual matrix elements of
the nonlocal Fock exchange operator together with the corresponding
KS energy denominators \cite{KummelKronik2008}. The resulting
$v_{x,\sigma}^{\mathrm{OEP}}$ is a local multiplicative potential
that yields the lowest exchange-only energy attainable within the
KS framework under the constraint of a common local one-electron
potential. It is free from one-electron self-interaction, exhibits
the correct $-1/r$ asymptotic behavior for finite systems, and
provides physically meaningful KS orbital energies
\cite{IvanovHirataBartlett1999,Goerling1999}.

Highly accurate numerical solutions of the atomic OEP equations were provided by Talman and Shadwick\cite{TalmanShadwick1976} and subsequently refined by Engel and co-workers \cite{EngelVosko1993,EngelHockDreizler2000,JiangEngel2005}. By exploiting spherical symmetry and representing the orbitals and potential on radial numerical grids, these approaches avoid an explicit auxiliary expansion of the exchange potential. They have produced benchmark-quality OEPx potentials for atoms and revealed characteristic features such as shell-dependent steps, intershell peaks, the correct near-nuclear behavior, and the asymptotic Coulombic tail. These numerical solutions remain valuable references for assessing approximate exchange potentials and finite-basis OEP implementations.


For molecules, the OEP equations are therefore usually solved in a finite orbital basis together with a separate auxiliary representation of the local potential \cite{IvanovHirataBartlett1999,Goerling1999,IvanovHirataBartlett2002}. A particularly useful starting point is the exact decomposition
\begin{equation}
v_{x,\sigma}^{\mathrm{OEP}}(\mathbf r)
=
v_{x,\sigma}^{\mathrm{Slater}}(\mathbf r)
+
v_{x,\sigma}^{\mathrm{resp}}(\mathbf r) + C
\label{eq:oep_decomposition}
\end{equation}
where $v_{x,\sigma}^{\mathrm{Slater}}$ is the Slater or exchange-hole potential contribution, and $v_{x,\sigma}^{\mathrm{resp}}$ contains the response of the exchange hole and the orbitals to a variation of the density, and $C$ is an additive constant. Introducing the spin-resolved one-particle density matrix $
\gamma_{\sigma}(\mathbf r,\mathbf r')
=
\sum_i^{\mathrm{occ}}
\phi_{i\sigma}(\mathbf r)
\phi_{i\sigma}^{*}(\mathbf r')
$
and $\rho_\sigma(\mathbf r)=\gamma_\sigma(\mathbf r,\mathbf r)$, the Slater potential can be written as
\begin{equation}
v_{x,\sigma}^{\mathrm{Slater}}(\mathbf r)
=-\frac{1}{\rho_\sigma(\mathbf r)}
\int
\frac{
|\gamma_\sigma(\mathbf r,\mathbf r')|^2
}{
|\mathbf r-\mathbf r'|
}
,d\mathbf r'.
\end{equation}
It is the electrostatic potential generated by the spin-resolved exchange hole. The Slater term contains the dominant attractive contribution to the exact exchange potential and already recovers its leading $-1/r$ asymptotic behavior. Nevertheless, it lacks most of the shell-dependent step structure of OEPx and systematically misses the remaining, predominantly positive, response contribution.

In a finite-basis implementation, the unknown response part is
typically represented as
\begin{equation}
v_{x,\sigma}^{\mathrm{resp}}(\mathbf r)
=
\sum_{t=1}^{N_{\mathrm{aux}}}
c_{t\sigma} f_t(\mathbf r),
\label{eq:response_expansion}
\end{equation}
where $\{f_t\}$ denotes a set of auxiliary basis functions. The OEP
integral equation is then projected onto the auxiliary space spanned
by these functions. The specific choice of auxiliary space depends
on the particular implementation. In the approach introduced by
Ivanov \etal, the OEP equation is projected directly onto an
auxiliary Gaussian space constructed from atomic-orbital (AO)
functions \cite{IvanovHirataBartlett1999,IvanovHirataBartlett2002}.
In Görling's formulation, an auxiliary exchange-charge density is
expanded instead, and the corresponding potential basis functions
are obtained as Coulomb transforms of Gaussian- or Slater-type charge
distributions \cite{Goerling1999,Gritsenko1996}. This construction provides a
natural route to the correct asymptotic behavior by imposing the
appropriate normalization of the exchange charge. Other
implementations employ dedicated Slater-type auxiliary functions,
whose radial form can be advantageous for representing both the
near-nuclear and asymptotic regions \cite{OEPslater}.

Despite their formal simplicity, finite-basis OEP calculations remain numerically delicate. The KS response operator possesses a constant null mode, reflecting the fact that the potential is defined only up to an additive constant\cite{hirata2001}. An auxiliary basis that is too small cannot reproduce the relevant structure of the potential, whereas an excessively flexible or unbalanced basis may generate strongly oscillatory and nonunique solutions\cite{balanceOEP}. The inversion of the projected response matrix consequently requires regularization, most commonly through a truncated singular-value decomposition (TSVD)\cite{SINGH2023297}, and the resulting potential can depend appreciably on the auxiliary basis and the chosen truncation threshold. 

These numerical difficulties have motivated the development of numerous approximations to the OEPx potential, including the Krieger-Li-Iafrate (KLI)\cite{KriegerLiIafrate1992}, localized HF (LHF)\cite{DellaSalaGoerling2001}, common-energy-denominator, Becke--Johnson (BJ)\cite{BeckeJohnson2006}, Räsänen--Pittalis--Proetto (RPP)\cite{RasanenPittalisProetto2010}, and Gritsenko--van Leeuwen--van Lenthe--Baerends (GLLB) constructions\cite{GritsenkoEtAl1995}. Although they follow different derivations and are not necessarily functional derivatives of an exchange-energy functional, they can all be cast in the structural form of Eq.~\eqref{eq:oep_decomposition}.

For example, using the positive kinetic-energy density, $\tau_\sigma(\mathbf r)
=
\frac{1}{2}
\sum_i^{\mathrm{occ}}
|\nabla\phi_{i\sigma}(\mathbf r)|^2,$
the BJ response potential is
\begin{equation}
v_{x,\sigma}^{\mathrm{resp,BJ}}(\mathbf r)
=
C_{\mathrm{BJ}}
\sqrt{
\frac{\tau_\sigma(\mathbf r)}
{\rho_\sigma(\mathbf r)}
},
\qquad
C_{\mathrm{BJ}}
=
\frac{1}{\pi}\sqrt{\frac{5}{6}}.
\label{eq:bj_response}
\end{equation}
where $C_{\mathrm{BJ}}$ is fixed by the homogeneous-electron-gas (HEG) limit. This simple semilocal term reproduces much of the atomic shell structure of OEPx. The RPP construction, in turn replaces $\tau_\sigma$ by its gauge-invariant Pauli-like component\cite{Smi-PRB-2020}
\begin{align}
v_{x,\sigma}^{\mathrm{resp,RPP}}(\mathbf r)
&=
C_{\mathrm{BJ}}
\sqrt{
\frac{D_\sigma(\mathbf r)}
{\rho_\sigma(\mathbf r)}
},
\label{eq:rpp_response}
\\
D_\sigma(\mathbf r)
&=
\tau_\sigma(\mathbf r)
-
\frac{|\nabla\rho_\sigma(\mathbf r)|^2}
     {8\rho_\sigma(\mathbf r)} \; .
\nonumber
\end{align}
This modification makes the correction gauge invariant, forces it to vanish for any one-electron density, and restores the appropriate asymptotic limit.

The GLLB response term is instead constructed from orbital-energy differences and normalized orbital densities,
\begin{equation}
v_{x,\sigma}^{\mathrm{resp,GLLB}}(\mathbf r)
=
C_x
\sum_i^{\mathrm{occ}}
\sqrt{\varepsilon_{H\sigma}-\varepsilon_{i\sigma}}\,
\frac{|\phi_{i\sigma}(\mathbf r)|^2}
     {\rho_\sigma(\mathbf r)},
\qquad
\\
C_x=\frac{8\sqrt{2}}{3\pi^2},
\label{eq:gllb_response}
\end{equation}
where $\varepsilon_{H\sigma}$ is the KS highest occupied molecular orbital (HOMO) energy. The square-root dependence follows from the uniform coordinate-scaling condition, whereas $C_x$ is fixed by the HEG limit. Because for HOMO the sum has a zero coefficient, the GLLB response vanishes asymptotically for finite systems.

Within KLI, the response potential is expanded directly in the normalized occupied-orbital densities,
\begin{align}
v_{x,\sigma}^{\mathrm{resp,KLI}}(\mathbf r)
=
\sum_i^{\mathrm{occ}}
w_{i\sigma}^{\mathrm{KLI}}
\frac{|\phi_{i\sigma}(\mathbf r)|^2}
     {\rho_\sigma(\mathbf r)},
\\
w_{i\sigma}^{\mathrm{KLI}} =
\left\langle
\phi_{i\sigma}
\left|
v_{x,\sigma}^{\mathrm{KLI}}
-\hat K_{\sigma}
\right|
\phi_{i\sigma}
\right\rangle ,
\label{eq:kli_response}
\end{align}
where $\hat K_{\sigma}$ is the nonlocal Fock exchange operator constructed from the KS orbitals. The additive constant is conventionally fixed by setting the coefficient associated with the HOMO to zero. LHF, which is equivalent to the self-consistent common-energy-denominator or effective-local-potential\cite{ELP} construction, extends the KLI form by retaining off-diagonal occupied--occupied contributions,
\begin{align}
v_{x,\sigma}^{\mathrm{resp,LHF}}(\mathbf r)
&=
\frac{1}{\rho_\sigma(\mathbf r)}
\sum_{i,j}^{\mathrm{occ}}
w_{ij,\sigma}^{\mathrm{LHF}}
\phi_{i\sigma}^{*}(\mathbf r)
\phi_{j\sigma}(\mathbf r),
\label{eq:lhf_response}
\\
w_{ij,\sigma}^{\mathrm{LHF}}
&=
\left\langle
\phi_{i\sigma}
\left|
v_{x,\sigma}^{\mathrm{LHF}}
-\hat K_{\sigma}
\right|
\phi_{j\sigma}
\right\rangle .
\nonumber
\end{align}
Thus, the KLI and LHF potentials are themselves finite expansions in physically meaningful orbital-density functions rather than in generic atom-centered auxiliary functions.

\begin{figure}[t]
\centering
\includegraphics[width=\columnwidth]{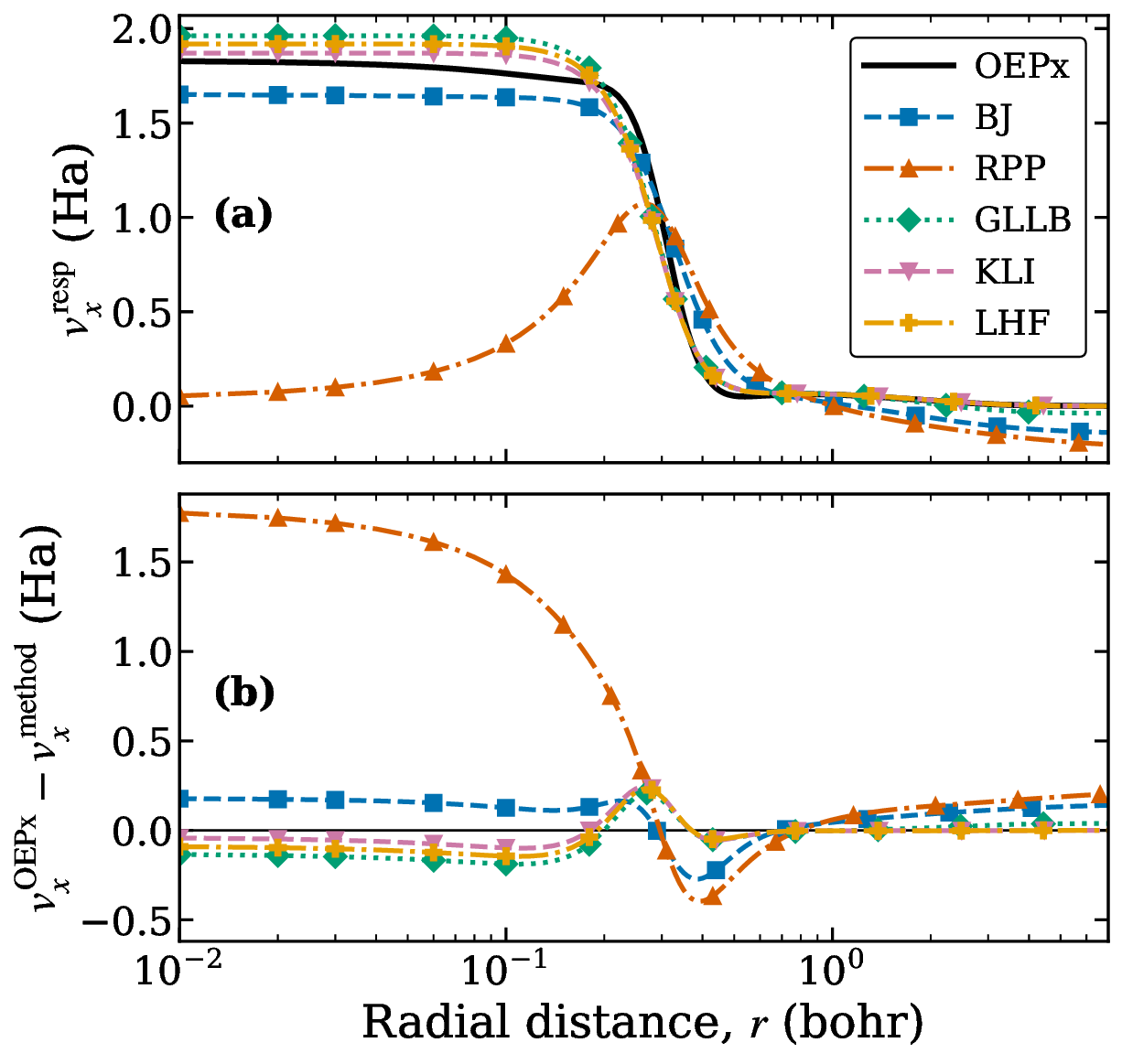}
\caption{
Comparison of approximate exchange-response corrections with the numerical OEPx reference for the Ne atom.
\textbf{(a)} Radial correction field,
$v_x^{\mathrm{OEPx}}(r)-v_x^{\mathrm{Slater}}(r)$,
together with the corresponding BJ, RPP, GLLB, KLI, and LHF approximations.
\textbf{(b)} Residual error of each approximation with respect to the full OEPx exchange potential,
$v_x^{\mathrm{OEPx}}(r)-v_x^{\mathrm{method}}(r)$,
where $\mathrm{method}=\mathrm{BJ}$, RPP, GLLB, KLI, or LHF.
The radial coordinate is shown on a logarithmic scale.
}
\label{fig:span_ne}
\end{figure}

\begin{figure}[t]
\centering
\includegraphics[width=\columnwidth]{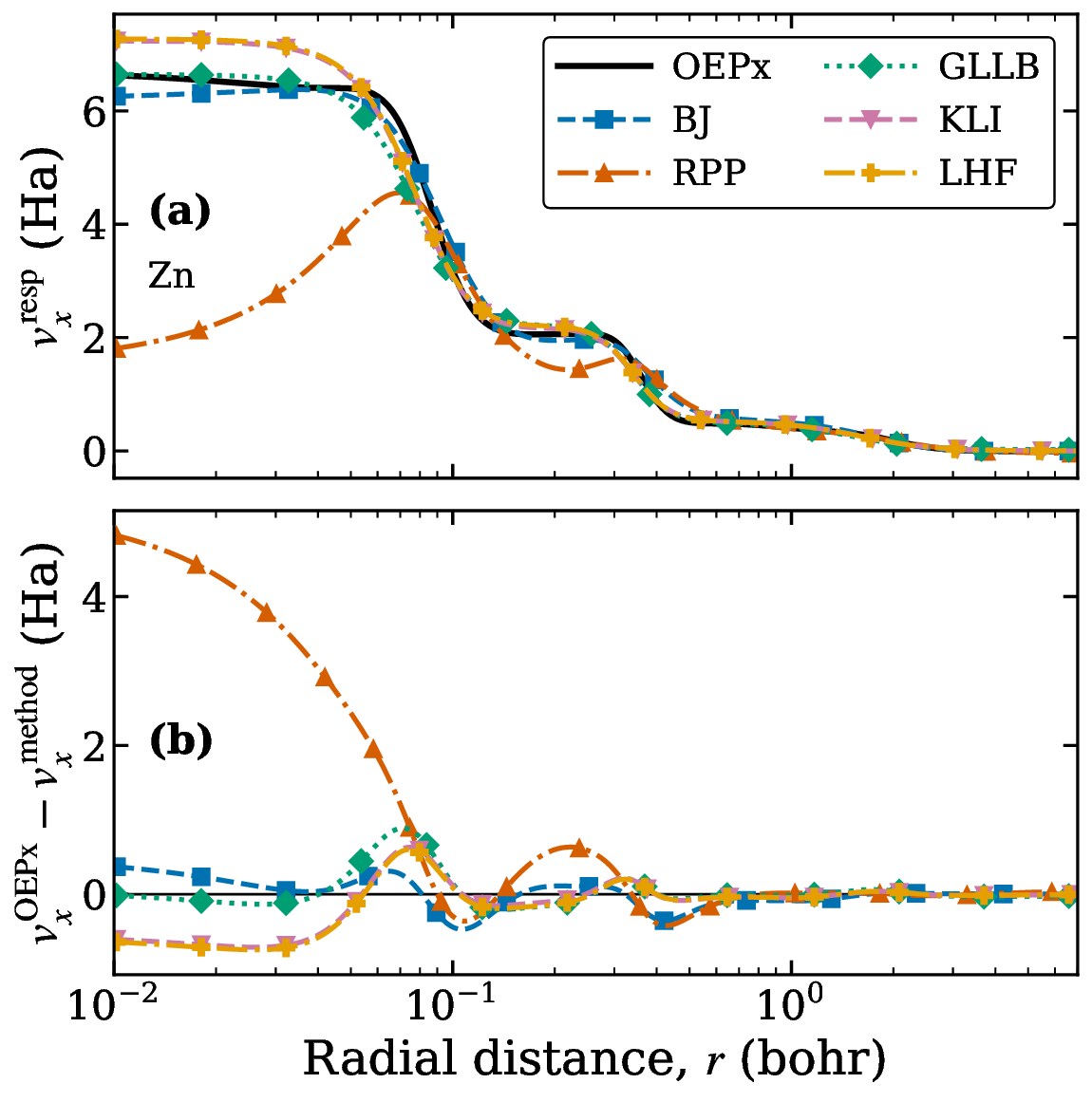}
\caption{
Comparison of approximate exchange-response corrections with the numerical OEPx reference for the Zn atom.
\textbf{(a)} Radial correction field,
$v_x^{\mathrm{OEPx}}(r)-v_x^{\mathrm{Slater}}(r)$,
together with the corresponding BJ, RPP, GLLB, KLI, and LHF approximations.
\textbf{(b)} Residual error of each approximation with respect to the full OEPx exchange potential,
$v_x^{\mathrm{OEPx}}(r)-v_x^{\mathrm{method}}(r)$,
where $\mathrm{method}=\mathrm{BJ}$, RPP, GLLB, KLI, or LHF.
The radial coordinate is shown on a logarithmic scale.
}
\label{fig:span_zn}
\end{figure}

\begin{figure}[t]
\centering
\includegraphics[width=\columnwidth]{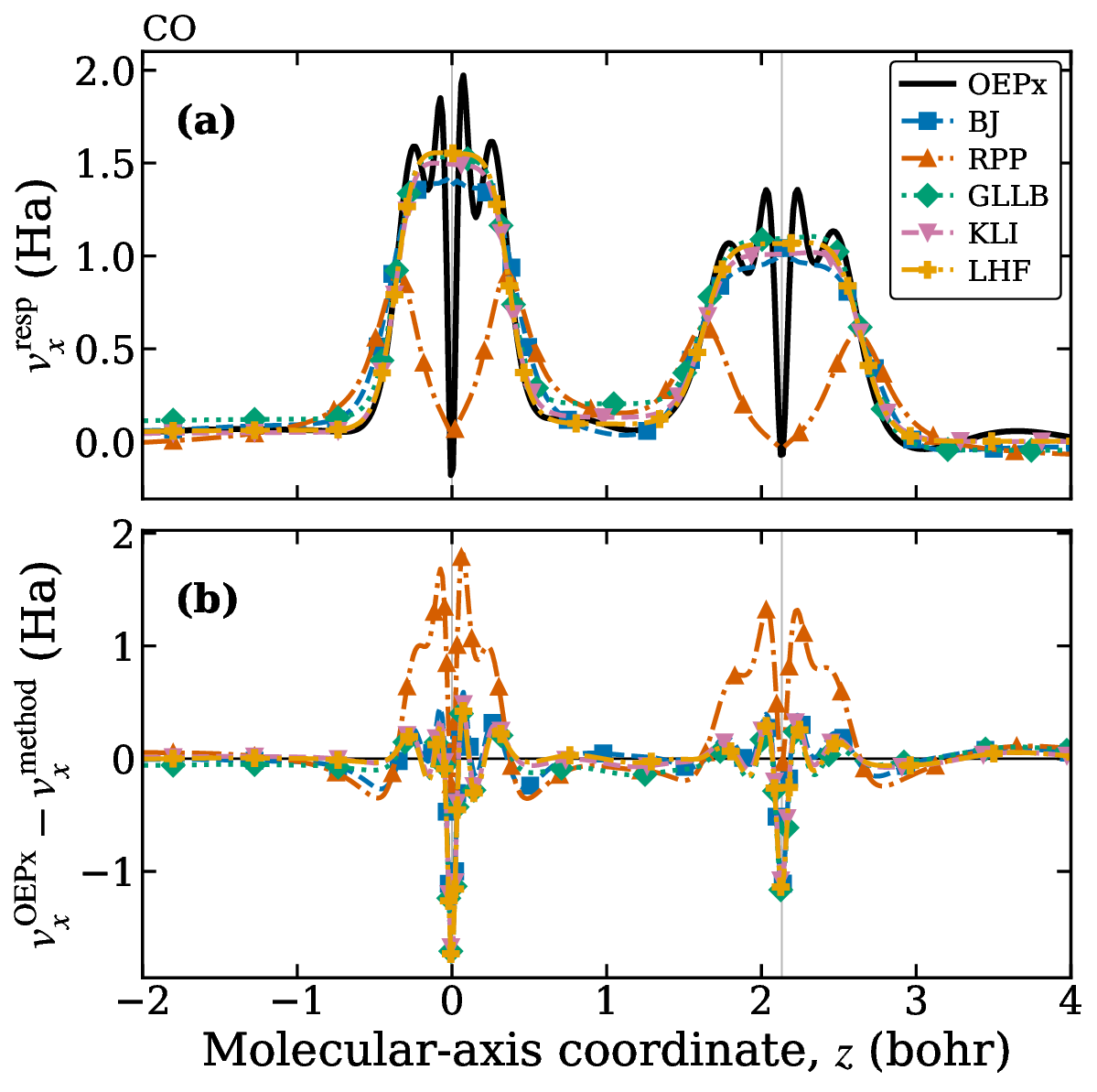}
\caption{
Comparison of approximate exchange-response corrections with the highly accurate OEPx reference data for the CO molecule.
\textbf{(a)} Radial correction field,
$v_x^{\mathrm{OEPx}}(r)-v_x^{\mathrm{Slater}}(r)$,
together with the corresponding BJ, RPP, GLLB, KLI, and LHF approximations.
\textbf{(b)} Residual error of each approximation with respect to the full OEPx exchange potential,
$v_x^{\mathrm{OEPx}}(r)-v_x^{\mathrm{method}}(r)$,
where $\mathrm{method}=\mathrm{BJ}$, RPP, GLLB, KLI, or LHF.
The calculations employed the aug-cc-pCVQZ orbital basis and the aug-cc-pwCVQZ-RIFIT auxiliary basis. 
}
\label{fig:span_co}
\end{figure}

The central observation motivating the present work is that these apparently different response approximations generate functions localized in the same chemically relevant spatial regions as the exact OEPx response. Figs.~\ref{fig:span_ne}, ~\ref{fig:span_zn}, and ~\ref{fig:span_co} illustrate this point for the Ne, Zn atoms and the CO molecule, respectively. In each case, the reference correction,
$v_x^{\mathrm{OEP}}-v_x^{\mathrm{Slater}}$, is compared with the BJ, RPP, and GLLB response terms and with the collective KLI- and LHF-type corrections. Although none of the individual approximations reproduces the OEPx response quantitatively over the entire spatial domain, they capture its dominant shell, intershell, bonding, and asymptotic features remarkably well. They therefore contain precisely the spatial information that a conventional auxiliary basis must recover through a much larger set of generic atom-centered functions.

This observation suggests an alternative strategy for solving or
approximating the OEP equations. Rather than representing the response
potential exclusively in terms of conventional AO-like auxiliary
functions, as in direct projection schemes, or Coulomb-weighted
auxiliary functions, as in exchange-charge formulations, one can use
the response factors underlying the BJ, RPP, GLLB, KLI, and LHF
approximations as a compact and physically motivated auxiliary space.
The coefficients associated with these response functions can then be
determined directly from the projected OEP equation, rather than being
fixed by the assumptions underlying the corresponding model
potentials. This construction retains the physical interpretation and
compact dimensionality of the underlying model potentials while
allowing their response amplitudes to adapt to the system-specific
OEP condition.

The purpose of this work is to demonstrate that model response functions can serve as an efficient auxiliary space for representing the OEPx correction to the Slater potential. On this basis, we introduce several adaptive exchange-potential approximations in which the amplitudes of selected BJ-, RPP-, GLLB-, KLI-, and LHF-inspired response factors are determined from the OEP condition. We examine their ability to reproduce the shape, asymptotic behavior, orbital energies, and exchange-only energetics of the full OEPx solution. The resulting methods systematically improve upon their original fixed-form counterparts and provide a compact connection between conventional finite-basis OEP algorithms and physically motivated approximate exchange potentials.

The rest of the paper is organized as follows. In Sec.~\ref{theory},
we introduce the theoretical framework for finite-basis OEPx and the
adaptive auxiliary response spaces, followed by a discussion of the
underlying model exchange potentials. Section~\ref{computation}
describes the computational details and the numerical implementation,
while Sec.~\ref{sec:results} presents and discusses the results.
Finally, in Sec.~\ref{sec:conclusions}, we summarize the main findings
and discuss possible directions for future work.

\section{Theory}
\label{theory}

\subsection{Finite-basis OEPx equations and potential gauge}

The response part in Eq.~\eqref{eq:oep_decomposition} is represented by auxiliary space expansion via Eq.~\eqref{eq:response_expansion}, where the response functions $f_p(\mathbf r)$ are specified below for each approximation. Spin indices are suppressed in the following discussion. At a fixed KS iteration, we define
\begin{equation}
d_{ia}^{p}=\langle i|f_p|a\rangle,\qquad
s_{ia}=\langle i|v_x^{\mathrm{Slater}}|a\rangle,\qquad
k_{ia}=\langle i|\hat K|a\rangle,
\end{equation}
where $i$ and $a$ label occupied and virtual orbitals, respectively. Projection of the OEP equation onto the chosen response space yields the following generic equations:
\begin{align}
X_{pq}
&=
2\sum_{ia}
\frac{d_{ia}^{p}d_{ia}^{q}}
     {\varepsilon_i-\varepsilon_a},
\label{eq:oep_projected_matrix}
\\
(x_{\mathrm{S}})_p
&=
2\sum_{ia}
\frac{d_{ia}^{p}s_{ia}}
     {\varepsilon_i-\varepsilon_a},
\label{eq:oep_slater_vector}
\\
\lambda_p
&=
-2\sum_{ia}
\frac{d_{ia}^{p}k_{ia}}
     {\varepsilon_i-\varepsilon_a},
\label{eq:oep_exchange_vector}
\end{align}
and hence
\begin{equation}
\mathbf X\mathbf c
=
\boldsymbol{\lambda}-\mathbf x_{\mathrm{S}}.
\label{eq:projected_oep}
\end{equation}
Equation~\eqref{eq:projected_oep} is the common working equation for all OEPx expansions considered in this work. Their only difference is the dimension and composition of the response space $\{f_p\}$. Equation.~\eqref{eq:projected_oep} is solved in every KS iteration using TSVD procedure\cite{SINGH2023297,hirata2001} pseudoinverse, discarding response eigenmodes satisfying
\begin{equation}
\frac{|\omega_q|}{|\omega_{\mathrm{max}}|}
\leq 10^{-6},
\label{eq:tsvd_threshold}
\end{equation}
where $\omega_q$ are the eigenvalues of the projected response matrix normalized by the largest value ($|\omega_{\max}|=\max_q|\omega_q|$). This procedure yields the minimum-norm solution in the linearly independent response space.

The OEP equation does not determine an additive constant in the local exchange potential (see \Eq{eq:oep_decomposition}). After the response coefficients have been obtained, the $C$ constant can be found by applying the HOMO condition~\cite{KummelKronik2008,IvanovHirataBartlett2002,SmigaIP}
\begin{equation}
C
=
\langle H|
-\hat K
-v_x
|H\rangle.
\label{eq:homo_shift}
\end{equation}
For degenerate HOMO shells, Eq.~\eqref{eq:homo_shift} is imposed on the shell average\cite{lhf2,IvanovHirataBartlett2002}.  The shift does not change the orbitals or total energy, but fixes the potential gauge and the absolute KS eigenvalue scale.


\subsection{Adaptive auxiliary response spaces}

\subsubsection{Adaptive BJ}

The adaptive BJ (aBJ) potential uses a one-dimensional ($N_{aux} = 1$) response space generated by
\begin{equation} 
f_1(\mathbf r) = f^{\mathrm{BJ}}(\mathbf r)
=
\sqrt{\frac{\tau(\mathbf r)}{\rho(\mathbf r)}}.
\label{eq:abj_basis}
\end{equation}
The potential is therefore
\begin{equation}
v_x^{\mathrm{aBJ}}(\mathbf r)
=
v_x^{\mathrm{Slater}}(\mathbf r)
+
c_{\mathrm{BJ}}f^{\mathrm{BJ}}(\mathbf r)
+
C_{\mathrm{aBJ}}.
\label{eq:abj_potential}
\end{equation}
The $c_{\mathrm{BJ}}$ coefficient follows directly from the scalar version of Eq.~\eqref{eq:projected_oep}
\begin{equation}
c_{\mathrm{BJ}}
=
\frac{\lambda_{\mathrm{BJ}}-(x_{\mathrm{S}})_{\mathrm{BJ}}}
     {X_{\mathrm{BJ},\mathrm{BJ}}}.
\label{eq:abj_coefficient}
\end{equation}
The original BJ form is recovered when $c_{\mathrm{BJ}}=\pi^{-1}\sqrt{5/6}$. Thus, aBJ retains the BJ response direction but replaces its universal HEG coefficient with the value selected by the projected OEP equation.

\subsubsection{Adaptive GLLB}

Similarly, we introduce the adaptive GLLB method (aGLLB), where the one-dimensional ($N_{aux} = 1$) response space is generated by the GLLB factor
\begin{equation}
f_1(\mathbf r) = f^{\mathrm{GLLB}}(\mathbf r)
=
\sum_i^{\mathrm{occ}}
\sqrt{\varepsilon_H-\varepsilon_i}\,
\frac{|\phi_i(\mathbf r)|^2}{\rho(\mathbf r)}.
\label{eq:agllb_basis}
\end{equation}
The corresponding potential,
\begin{equation}
v_x^{\mathrm{aGLLB}}(\mathbf r)
=
v_x^{\mathrm{Slater}}(\mathbf r)
+
c_{\mathrm{GLLB}}f^{\mathrm{GLLB}}(\mathbf r)
+
C_{\mathrm{aGLLB}},
\label{eq:agllb_potential}
\end{equation}
is obtained with
\begin{equation}
c_{\mathrm{GLLB}}
=
\frac{\lambda_{\mathrm{GLLB}}-(x_{\mathrm{S}})_{\mathrm{GLLB}}}
     {X_{\mathrm{GLLB},\mathrm{GLLB}}}.
\label{eq:agllb_coefficient}
\end{equation}
The conventional GLLB response is recovered for
$c_{\mathrm{GLLB}}=8\sqrt{2}/(3\pi^2)$. As in aBJ, the adaptive step amounts to solving a single scalar equation at every KS iteration.

\subsubsection{Adaptive RPP and GLLB}

As shown in \Fig{fig:span_ne} and \Fig{fig:span_co}, the RPP term behaves spatially differently than all others, possibly introducing more flexibility in the shell oscillation region. Thus, in order to test this possibility, we introduce the two-dimensional ($N_{aux} = 2$) response space consisting of RPP and GLLB response directions, denoted as the aRG method.

The aRG scheme combines the GLLB (\Eq{eq:agllb_basis}) and RPP response directions, additionally using
\begin{equation}
f_2(\mathbf r) = f^{\mathrm{RPP}}(\mathbf r)
=
\sqrt{\frac{D(\mathbf r)}{\rho(\mathbf r)}} \; .
\label{eq:rg_rpp_basis}
\end{equation}
The potential then reads 
\begin{equation}
v_x^{\mathrm{aRG}}(\mathbf r)
=
v_x^{\mathrm{Slater}}(\mathbf r) + 
c_{\mathrm{GLLB}}f^{\mathrm{GLLB}}(\mathbf r) \\
+
c_{\mathrm{RPP}}f^{\mathrm{RPP}}(\mathbf r)
+
C_{\mathrm{aRG}}\; ,
\label{eq:rg_potential}
\end{equation}
where the two coefficients are determined simultaneously from the aRG matrix equation
\begin{equation}
\scriptsize
\begin{pmatrix}
X_{\mathrm{GLLB,GLLB}} & X_{\mathrm{GLLB,RPP}}\\
X_{\mathrm{RPP,GLLB}} & X_{\mathrm{RPP,RPP}}
\end{pmatrix}
\begin{pmatrix}
c_{\mathrm{GLLB}}\\
c_{\mathrm{RPP}}
\end{pmatrix}
=
\begin{pmatrix}
\lambda_{\mathrm{GLLB}}-(x_{\mathrm S})_{\mathrm{GLLB}}\\
\lambda_{\mathrm{RPP}}-(x_{\mathrm S})_{\mathrm{RPP}}
\end{pmatrix}.
\normalsize
\label{eq:rg_equations}
\end{equation}
The adaptive solution determines the relative weights of the two complementary response fields from the OEP condition.

\subsubsection{Adaptive KLI}

The adaptive KLI (aKLI) scheme uses the KLI orbital-density ratios as an auxiliary basis set to span the response function
\begin{equation}
f_i^{\mathrm{KLI}}(\mathbf r)
=
\frac{|\phi_i(\mathbf r)|^2}{\rho(\mathbf r)}.
\label{eq:aKLI_basis}
\end{equation}
The aKLI potential is therefore
\begin{equation}
v_x^{\mathrm{aKLI}}(\mathbf r)
=
v_x^{\mathrm{Slater}}(\mathbf r)
+
\sum_{i\ne H}^{\mathrm{occ}}
c_i^{\mathrm{aKLI}}f_i^{\mathrm{KLI}}(\mathbf r)
+
C_{\mathrm{aKLI}}.
\label{eq:aKLI_potential}
\end{equation}
Because $\sum_i^{\mathrm{occ}}f_i^{\mathrm{KLI}}(\mathbf r)=1$, one diagonal function (or several functions in the case of a degenerate HOMO), chosen as the HOMO contribution, is omitted from the response space and represented by the separate constant in Eq.~\eqref{eq:aKLI_potential}. 
The coefficients are found from Eq.~\eqref{eq:projected_oep} in an $N_{aux} = (N_{\mathrm{occ}}-N_H)$ dimensional response space, where $N_H$ is the number of degenerate HOMO orbitals.  Although this space is identical to that used by the conventional KLI potential, aKLI is not obtained from the KLI coefficient equations. Instead, it is the full OEP equation projected onto the KLI response space. In particular, the coefficients are determined by the occupied-virtual response kernel in Eqs.~\eqref{eq:oep_projected_matrix}--\eqref{eq:oep_exchange_vector}.

\subsubsection{Adaptive LHF}

The adaptive LHF (aLHF) scheme extends aKLI by including the off-diagonal occupied-orbital products that arise in the LHF potential
\begin{equation}
f_{ij}^{\mathrm{LHF}}(\mathbf r)
=
\begin{cases}
\dfrac{|\phi_i(\mathbf r)|^2}{\rho(\mathbf r)}, & i=j,\\[8pt]
\dfrac{2\phi_i(\mathbf r)\phi_j(\mathbf r)}{\rho(\mathbf r)}, & i<j.
\end{cases}
\label{eq:aLHF_basis}
\end{equation}
Only products compatible with the symmetry of the local potential are retained.  As for aKLI, one diagonal (or multiple in the case of degenerate HOMO state) function is removed because the sum of all diagonal terms is constant.  The aLHF potential becomes
\begin{equation}
v_x^{\mathrm{aLHF}}(\mathbf r)
=
v_x^{\mathrm{Slater}}(\mathbf r)
+
\sum_{(ij)\in\mathcal I}
c_{ij}^{\mathrm{aLHF}}f_{ij}^{\mathrm{LHF}}(\mathbf r)
+
C_{\mathrm{aLHF}},
\label{eq:aLHF_potential}
\end{equation}
where $\mathcal I$ is the set of retained occupied--occupied products. 
Before symmetry reduction, the dimension of this space is $N_{aux} = \left(N_{\mathrm{occ}}(N_{\mathrm{occ}}+1)\right)/2-N_H$.

The aLHF differs from conventional LHF in the same way that aKLI differs from KLI: the basis functions are LHF-like, but the coefficients are determined by projection of the complete OEP equation rather than from the conventional LHF self-consistency condition.


\section{Computational details}
\label{computation}

All calculations have been performed with a locally modified version of {\tt PySCF}~\cite{pyscf} program. As in our previous studies\cite{smiga2016self,smiga2019self,grabowski:2011:jcp,SmigaIP,OEPSOSszs,ISIOEP} in order to solve OEP equation (\Eq{eq:projected_oep}) we have employed the
finite-basis set procedure of \RRef{IvanovHirataBartlett1999}.
To calculate the pseudo-inverse of the density-density response matrix, we have utilized a TSVD with the cutoff $10^{-6}$ (see \Eq{eq:tsvd_threshold}). This step is essential for determining stable and physically meaningful OEP solutions \cite{hirata2001,IvanovHirataBartlett2002,OEPSOSszs, SINGH2023297}.

The calculations have been performed for the 16 closed-shell systems considered in our previous works~\cite{OEPSOSszs,SCF-ISI}. In all calculations we employed uncontracted triple-zeta-quality basis sets as in \RRef{OEPSOSszs}, namely an even tempered $20s10p2d$ basis set \cite{grabowski:2011:jcp} for He and He$_2$, an uncontracted ROOS-ATZP basis set~\cite{widm90} for Be, the Ne atom and Ne$_2$. For the Ar atom, we used a modified basis set which combines $s$ and $p$ type basis functions 
from the uncontracted ROOS--ATZP~\cite{widm90} with $d$ and $f$ functions coming from the uncontracted aug--cc--pwCVQZ basis set~\cite{peterson02}. The remaining systems were treated in the uncontracted cc-pVTZ basis set of Dunning~\cite{dunning:1989:bas}. For all molecular systems, we considered their equilibrium geometries as in \RRef{OEPSOSszs}.

The Slater potential in \Eq{eq:oep_decomposition} was evaluated using the density-fitting (DF) procedure described in \RRef{DellaSalaGoerling2001} using cc-pVTZ-JKFIT basis set for the cc-pVTZ orbital bases, aug-cc-pVTZ-JKFIT for Ar, and automatically generated even-tempered auxiliary bases for He, He$_2$, Be, Ne, Ne$_2$, and Mg available in {\tt PySCF}. We note that, for a few tested cases (not reported), the full-Slater and DF Slater implementations give effectively identical self-consistent results. This is due to the fact that adaptive response term compensates most of this difference. Therefore, DF Slater is a numerically safe and much faster guiding potential for these calculations.


At every self-consistent iteration, all matrix elements (\Eq{eq:oep_projected_matrix}-\Eq{eq:oep_exchange_vector}) and DF Slater potential
were rebuilt from the
current orbitals. In all calculations, we used the HF-converged orbitals as the initial guess.
The convergence criterion was set to $10^{-8}$ for the maximum density-matrix change.

Figures~\ref{fig:span_ne} and~\ref{fig:span_zn} were generated from the
numerical atomic OEPx code of Engel and co-workers\cite{EngelVosko1993}.  Figure~\ref{fig:span_co}
was generated with our in-house PySCF OEP implementation which will be published in separate work. As stated in its caption, that CO calculation used the uncontracted aug-cc-pCVQZ orbital basis
and aug-cc-pwCVQZ-RIFIT auxiliary basis. All geometries and the data supporting this finding are available in an external repository~\cite{adaptive2026}.

\section{Results}
\label{sec:results}

\subsection{Total energies and HOMO ionization-potentials}

\begin{figure*}[t]
\centering
\includegraphics[width=0.9\textwidth]{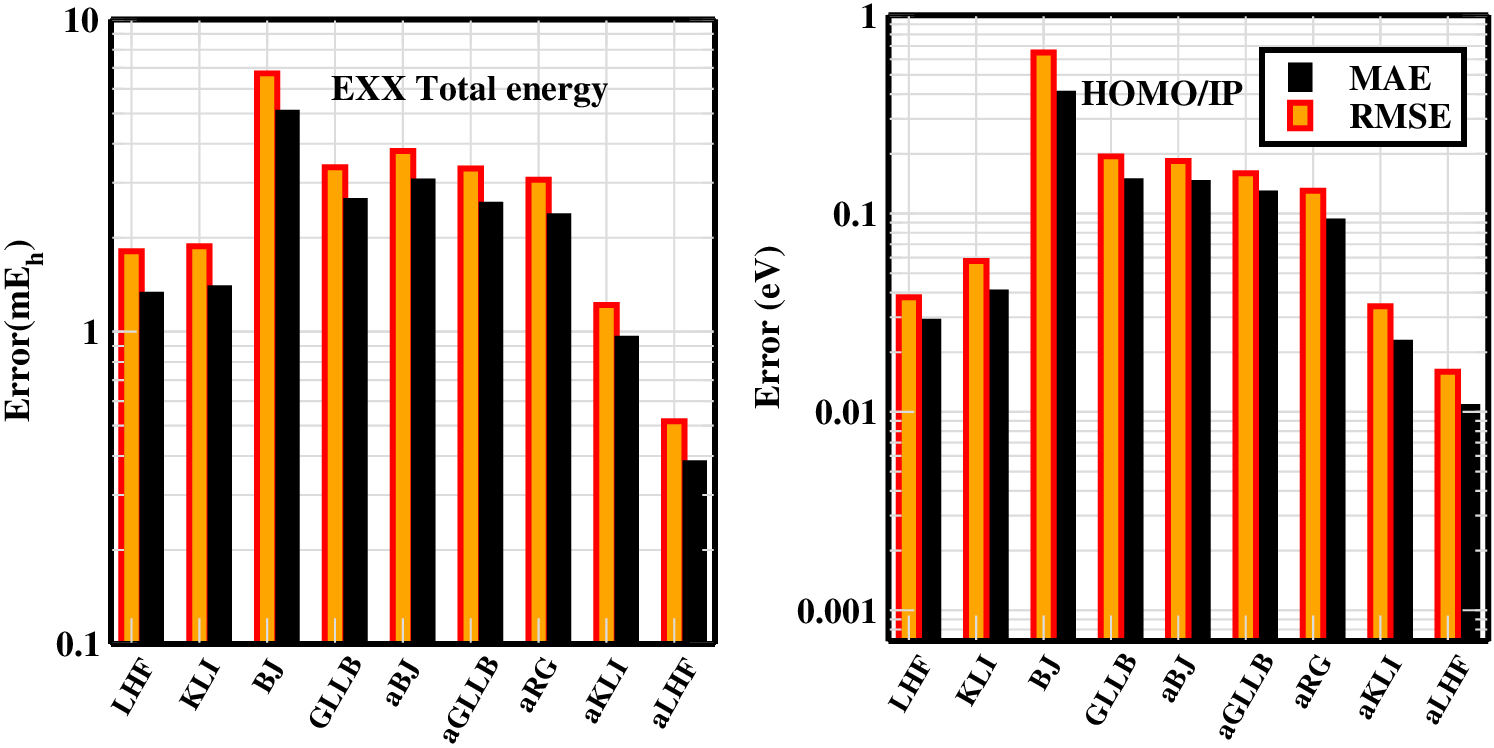}
\caption{MAE and RMSE for the exchange-only total energy (left) and HOMO/IP (right) over the 16-system benchmark.  All errors are relative to the self-consistent finite-basis OEPx calculation. }
\label{fig:errors}
\end{figure*}

Table~\ref{tab:benchmark_errors} summarizes the deviations of the approximate and adaptive
potentials from the self-consistent finite-basis OEPx results\cite{IvanovHirataBartlett1999} . We
consider the total-energy and HOMO ionization-potential (IP) errors,
$I=-\varepsilon_H$, evaluated in terms of the mean absolute error
(MAE), root-mean-square error (RMSE), and mean absolute relative error
(MARE) over the 16-system benchmark set. The MAE and RMSE values are
additionally visualized in \Fig{fig:errors}.

Among the conventional approximations, LHF and KLI show the smallest
deviations from OEPx, with total-energy MAEs of 1.325 and
1.389~m$E_h$, respectively, and HOMO IP MAEs of
0.0288 and 0.0404~eV. Consistent with the analysis of \RRef{grabowski:2013:molphys}, these
results again demonstrate the close relationship between the KLI and
LHF constructions. The largest deviations are obtained with the BJ
approximation, whose total-energy and HOMO IP MAEs
reach 5.066~m$E_h$ and 0.406~eV, respectively. This behavior can be
associated with inaccuracies in the representation of the BJ response
term, particularly in the core and asymptotic regions (see, e.g.,
\Fig{fig:span_ne}). GLLB performs better, with corresponding MAEs of
2.639~m$E_h$ and 0.147~eV, respectively, which is consistent with its
more appropriate representation of the response contribution compared
with BJ.

Adaptive optimization reduces the errors of both BJ and GLLB, although
the magnitude of the improvement differs substantially. For aBJ, the
total-energy MAE decreases from 5.066 to 3.054~m$E_h$, while the HOMO IP MAE decreases from 0.406 to 0.144~eV. Thus,
coefficient optimization reduces the respective errors by approximately
40\% and 65\%. For aGLLB, the improvement is more modest: the
total-energy MAE decreases from 2.639 to 2.571~m$E_h$, while the HOMO IP MAE decreases from 0.147 to 0.128~eV. This
difference is reflected in the optimized coefficients. For aBJ, the
mean $C_{\mathrm{BJ}}$ coefficient over the 16 systems is 0.24345,
which is lower than the original HEG value by 0.04713 (16.2\%).
Thus, the projected OEP equation systematically reduces the BJ amplitude on average, improves the representation of the core and
asymptotic regions of the potential. In contrast, aGLLB gives a mean
coefficient of 0.36959, only 0.01251 (3.3\%) below the HEG $C_x$ value.
The conventional GLLB amplitude is therefore already close to the mean
adaptive value, whereas the HEG BJ coefficient is, on average, too
large for the finite systems considered here.

The results further show that optimizing a single response coefficient
does not fully recover the OEP response. Adding the independent RPP
direction in the aRG model provides a further reduction in the errors,
yielding total-energy and HOMO IP MAEs of
2.357~m$E_h$ and 0.0922~eV, respectively. The mean RPP coefficient is
0.02055, corresponding to 93.0\% below its HEG value. This indicates
that the GLLB direction accounts for most of the response represented
in the aRG space, while the RPP direction provides a smaller but
independent correction to the spatial structure of the potential.
Relative to aGLLB, the addition of RPP reduces the total-energy MAE by
0.214~m$E_h$ and the HOMO IP MAE by 0.0354~eV.

A substantially larger improvement is obtained with the occupied-space
expansions. The aKLI and aLHF constructions give total-energy MAEs of
0.957 and 0.382~m$E_h$, respectively, and HOMO IP
MAEs of 0.0226 and 0.0107~eV. Thus, aLHF gives the smallest errors
among the response spaces considered here. The progressive improvement
from aGLLB to aRG, aKLI, and aLHF indicates that the accuracy of the
adaptive construction depends not only on optimizing the coefficients
but also on the dimensionality and physical content of the response
space. In particular, the results show that a compact one-dimensional
response direction can substantially improve upon a fixed model
potential, but additional physically motivated response directions
provide further flexibility to reproduce the finite-basis OEP response.

Overall, the results reveal a clear accuracy-dimensionality trade-off.
The adaptive models retain the compactness of physically motivated
response spaces while allowing their amplitudes to be determined
directly from the projected OEP equation. Increasing the dimensionality
of the response space progressively reduces the errors, with the aLHF
occupied-pair expansion providing the closest agreement with OEPx
among the spaces examined here.

\begin{table}[t]
\caption{Mean absolute error (MAE), root-mean-square error (RMSE), and mean absolute relative error (MARE) over the 16-system exchange-only benchmark. Energy errors are relative to the OEPx total energy; HOMO errors are identical to errors in $I=-\varepsilon_H$. The complete per-system values can be found in \RRef{adaptive2026}.}
\label{tab:benchmark_errors}
\centering
\scriptsize
\setlength{\tabcolsep}{3pt}
\renewcommand{\arraystretch}{1.18}
\begin{tabular}{@{}lccc|ccc@{}}
\toprule
\hline
\\[-7pt]
& \multicolumn{3}{c|}{Total energy}
& \multicolumn{3}{c}{$I=-\varepsilon_H$} \\
\hline
\cmidrule(lr){2-4}\cmidrule(l){5-7}
Method & MAE & RMSE & MARE (\%)
       & MAE & RMSE & MARE (\%) \\[-1pt]
\hline
& \multicolumn{2}{c}{m$E_h$} &
& \multicolumn{2}{c}{eV} & \\
\midrule
\hline
LHF   & 1.325 & 1.809 & 0.000919 & 0.0288 & 0.0379 & 0.196 \\
KLI   & 1.389 & 1.880 & 0.000976 & 0.0404 & 0.0577 & 0.285 \\
BJ    & 5.066 & 6.725 & 0.011411 & 0.4057 & 0.6473 & 2.198 \\
GLLB  & 2.639 & 3.362 & 0.002186 & 0.1472 & 0.1939 & 1.043 \\
aBJ   & 3.054 & 3.793 & 0.002468 & 0.1439 & 0.1840 & 0.986 \\
aGLLB & 2.571 & 3.336 & 0.001963 & 0.1276 & 0.1596 & 0.855 \\
aRG  & 2.357 & 3.066 & 0.001760 & 0.0922 & 0.1303 & 0.645 \\
aKLI  & 0.957 & 1.219 & 0.000725 & 0.0226 & 0.0341 & 0.167 \\
aLHF  & 0.382 & 0.517 & 0.000426 & 0.0107 & 0.0159 & 0.072 \\
\bottomrule
\hline
\end{tabular}
\end{table}


\subsection{Response spectra and numerical conditioning}

\begin{figure}[t]
\centering
\includegraphics[width=\columnwidth]{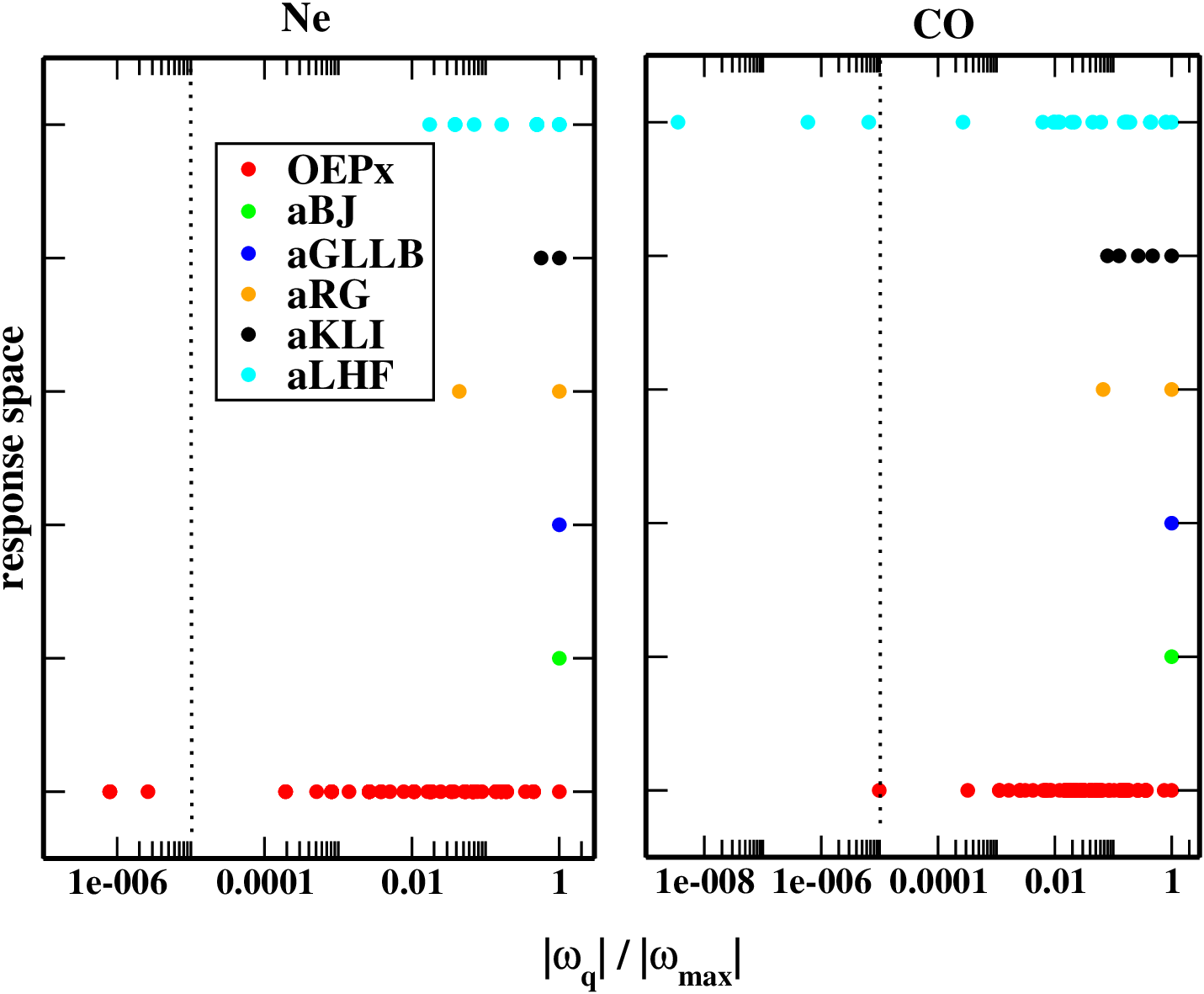}
\caption{Normalized eigenvalue magnitudes of the symmetrized projected OEP response matrices for Ne and CO.  The dashed vertical line marks the relative TSVD cutoff of $10^{-6}$.  The exact constant gauge has been removed; the remaining aLHF near-null modes are retained in the data file with their signed eigenvalues and are excluded from the minimum-norm solution when they fall below the cutoff.}
\label{fig:response}
\end{figure}

Figure~\ref{fig:response}  shows the absolute eigenvalue spectra of the
symmetrized projected OEP density-density response matrices for two representative cases, namely Ne atom and CO molecule.
For each system and response space, the eigenvalue magnitudes
are normalized by the largest value $|\omega_{\max}|$. After removal of the exact
constant-potential gauge mode, the TSVD pseudoinverse retains
only the set of response modes
\[
 \mathcal{R}
 =
 \left\{
 q:
 \frac{|\omega_q|}{|\omega_{\max}|}>10^{-6}
 \right\}.
\]
Accordingly, the dimension reported in Table~\ref{tab:response_spaces} is the nominal
number of auxiliary response directions entering the projected
problem, whereas the retained rank is
$r_{\mathrm{TSVD}}=|\mathcal{R}|$, i.e., the number of modes
actually used to construct the minimum-norm solution.

The condition number reported in Table~\ref{tab:response_spaces} refers exclusively to
this retained response space and is defined as $\kappa =|\omega_{\max}| /\displaystyle\min_{q\in\mathcal{R}}|\omega_q|$. 
Thus, $\kappa$ measures the spectral range that is
still inverted after TSVD regularization. 
Moreover, any
nonzero one-dimensional response space has
$\kappa=1$ identically. Consequently, the unit
condition numbers of aBJ and aGLLB indicate the absence of
internal linear dependence, but do not by themselves establish
that the corresponding response direction has a physically
significant magnitude.

For Ne, the conventional finite-basis OEPx auxiliary space
contains 82 directions, of which 79 survive the TSVD cutoff. Its
retained condition number, $\kappa=3.82\times10^{5}$, corresponds to a
smallest retained relative eigenvalue of approximately
$2.62\times10^{-6}$. The three excluded directions therefore
belong to the numerically unresolved part of the response space.
In contrast, all modes of the adaptive spaces are retained:
aBJ and aGLLB contain one direction each, aRG and aKLI contain
two directions, and aLHF contains 14 directions. The aRG and
aKLI spaces remain well conditioned, with
$\kappa=22.8$ and $1.77$, respectively. Despite
its substantially greater flexibility, the 14-dimensional aLHF
space also remains numerically stable for Ne, with
$\kappa=57.4$.

For CO, the conventional OEPx auxiliary space contains 84
directions and all of them remain above the chosen threshold.
Nevertheless, its condition number of
$1.01\times10^{5}$, corresponding to a smallest relative
eigenvalue of approximately $9.90\times10^{-6}$, shows that the
response spectrum already spans five orders of magnitude. The
compact aRG and aKLI spaces have dimensions two and five and
condition numbers of 14.9 and 12.5, respectively. The
occupied-pair aLHF construction increases the nominal dimension
to 27. In this case, two near-null directions fall below the
TSVD threshold, leaving a retained rank of 25 and
$\kappa=1.54\times10^{5}$. The smallest retained
aLHF eigenvalue is therefore only about
$6.49\times10^{-6}$ of the largest one. This result illustrates
that the additional flexibility of the occupied-pair expansion
is accompanied by weakly identifiable combinations of response
functions, particularly for molecular systems.

The role of TSVD is therefore not merely technical\cite{SINGH2023297}. Inverting a
near-null response mode would amplify small numerical errors in
the projected right-hand side, as well as changes in the
orbitals between self-consistent iterations, producing large and
poorly determined coefficients and potentially oscillatory
potentials. TSVD removes these unresolved directions and selects
the minimum-norm solution in the identifiable subspace. At the
same time, regularization cannot compensate for a response space
that lacks the relevant physical structure. This distinction is
particularly important for rank-one models. For example, in the He$_2$ system the GLLB response becomes very
small, so that its coefficient is effectively non-identifiable
even though the formal condition number of the one-dimensional
space is unity.

The spectra therefore expose a three-way balance between
dimensionality, physical flexibility, and numerical stability.
The low-dimensional aRG and aKLI spaces avoid the near-linear
dependencies characteristic of generic auxiliary expansions,
whereas aLHF approaches the flexibility and accuracy of the full
OEPx space at the cost of a broader spectrum and an increasing
need for TSVD regularization. 

\begin{table}
\caption{Dimensions, retained ranks, and retained response-space condition numbers for the OEP-projected spaces used in Fig.~\ref{fig:response}.  The OEPx values refer to the finite auxiliary OEP space; the other rows refer to the compact adaptive spaces. }
\label{tab:response_spaces}
\centering
\scriptsize
\begin{tabular}{lrrr|rrr}
\toprule
\hline\hline
 & \multicolumn{3}{c|}{Ne} & \multicolumn{3}{c}{CO}\\
 \hline
Method & dim. & rank & $\kappa$ & dim. & rank & $\kappa$\\
\hline
\midrule
OEPx  & 82 & 79 & $3.82\times10^5$ & 84 & 84 & $1.01\times10^5$\\
aBJ   & 1  & 1  & 1.00 & 1  & 1  & 1.00\\
aGLLB & 1  & 1  & 1.00 & 1  & 1  & 1.00\\
aRG  & 2  & 2  & 22.8 & 2  & 2  & 14.9\\
aKLI  & 2  & 2  & 1.77 & 5  & 5  & 12.5\\
aLHF  & 14 & 14 & 57.4 & 27 & 25 & $1.54\times10^5$\\
\hline
\bottomrule
\end{tabular}
\end{table}

\subsection{Potentials and densities}

\begin{figure}[t]
\centering
\includegraphics[width=\columnwidth]{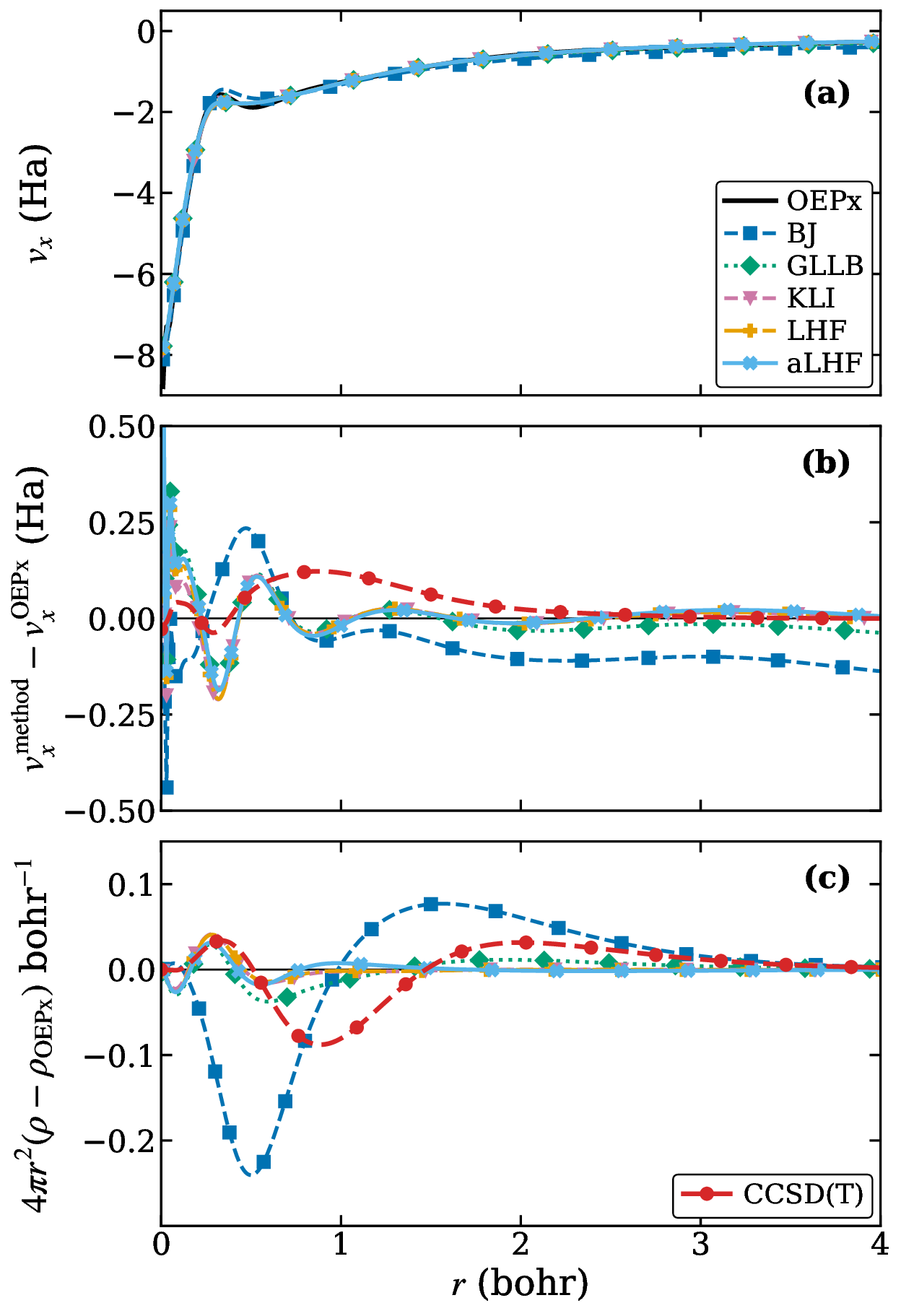}
\caption{Self-consistent Ne exchange potentials and densities. The upper panel
shows the exchange potentials $v_x$ obtained from OEPx and the approximate
exchange-only potentials considered in this work. The middle panel shows the
deviation of each exchange-only potential from the OEPx reference. The lower panel shows the associated density difference. At the middle and lower panel the CCSD(T) correlation potential and densities are placed for comparison.}

\label{fig:ne_results}
\end{figure}

\begin{figure}[t]
\centering
\includegraphics[width=\columnwidth]{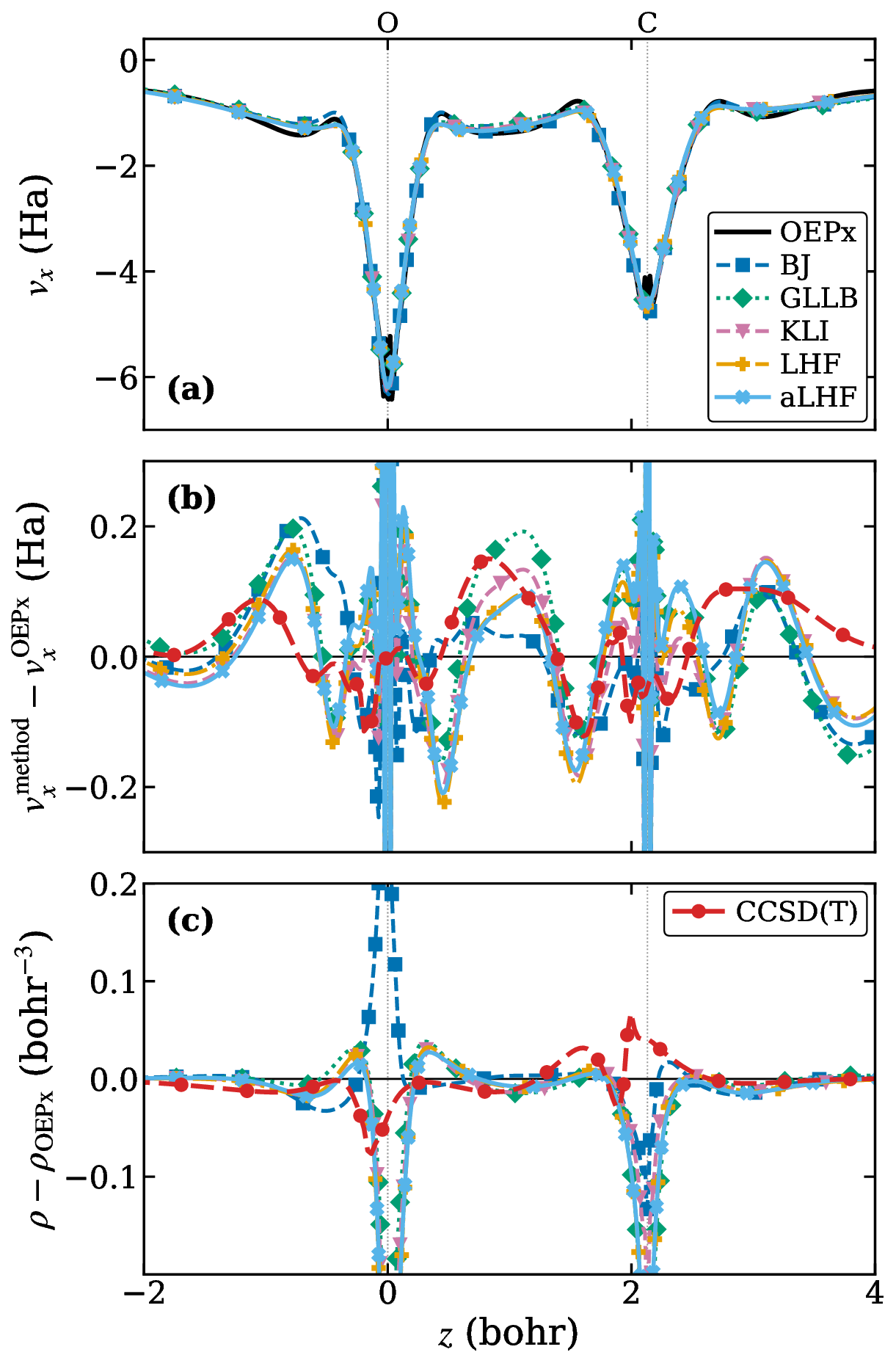}
\caption{Self-consistent CO exchange potentials and densities along the molecular
axis. The upper panel shows
the total exchange potentials $v_x$ obtained from OEPx and the approximate
exchange-only potentials. The middle panel shows the corresponding deviations
from OEPx. The lower panel shows the corresponding density
differences. At the middle and lower panel the CCSD(T) correlation potential and densities are placed for comparison.}

\label{fig:co_results}
\end{figure}

The spatial behavior of the self-consistent exchange potentials for Ne
and CO is shown in Figs.~\ref{fig:ne_results} and~\ref{fig:co_results}, respectively.  The upper and middle panels compare
the exchange potentials and their deviations from the OEPx reference ($v_x^{\mathrm{method}}(r)-v_x^{\mathrm{OEPx}}(r)$),
while the lower panels show the corresponding density differences. We
focus on aLHF together with the conventional BJ, GLLB, KLI, and LHF
forms to illustrate the progression from simple model potentials to
the adaptive response-based description.

For Ne, all potentials reproduce the overall shell structure of $v_x$ and
approach similar gauge-aligned tails.
The deviations from OEPx are therefore concentrated mainly in
the valence and intershell regions. The aLHF potential follows the OEPx
reference closely over most of the radial range, while BJ and GLLB show
larger deviations. The corresponding density differences are also
smaller for aLHF, consistent with its lower energetic errors in
Table~\ref{tab:benchmark_errors}.

A similar behavior is observed for CO, although the molecular
structure introduces additional spatial features near the two nuclear
cusps. Away from the nuclei, aLHF closely follows the OEPx potential,
whereas BJ, GLLB, and the other low-dimensional approximations exhibit
larger deviations and more pronounced oscillatory structure. The
corresponding density differences are localized mainly in the
core-valence and bonding regions. Thus, the adaptive response
construction improves not only the global energetic measures but also
the spatial representation of the local exchange potential and the
resulting self-consistent density. These observations are in line with those observed in \RRef{grabowski:2013:molphys}.

The comparison may also be viewed in the context of recent analyses of correlation effects in hybrid functionals~\cite{UPBH,vignesh2026}.  These studies showed that semilocal and exact-exchange contributions can reproduce important spatial features of correlated potentials and densities, including structures observed in CCSD(T) based references. Accordingly, the middle and bottom panels of Figs.~\ref{fig:ne_results} and~\ref{fig:co_results} include the available CCSD(T) reference data.  This comparison allows one to examine whether the residual exchange-potential errors possess spatial components similar to those associated with correlation.

In several regions, the approximate response terms indeed produce residual features with the same sign and shell or bond structure as the CCSD(T) correlation potential.  This is particularly evident for CO, where the oscillatory profile ($v_x^{\mathrm{method}}(r)-v_x^{\mathrm{OEPx}}(r)$) resembles the shell structure of the correlation potential over a substantial part of the molecular axis.  This resemblance should be interpreted qualitatively i.e. the exchange-only approximations do not explicitly include correlation, but their response functions may span spatial directions that are also important in the correlated potential.

At the density level, the closest agreement with the CCSD(T) reference for Ne occurs in the core and valence regions, consistent with \RRef{grabowski:2013:molphys}.  In the remaining radial regions, all methods except BJ remain close to the OEPx density. For CO, a similar pattern is found in the bond and tail regions.  Near the nuclei, however, sizeable density-difference peaks remain, as is typical of semilocal potential approximations\cite{grabowski:2013:molphys}.  These results suggest that part of the apparent success of BJ- and GLLB-based response models may result from cancellation between residual exchange-potential errors and missing correlation effects, analogous to the error-cancellation mechanisms discussed for semilocal and hybrid approximations~\cite{GLLBSL,GLLBS,UPBH,vignesh2026}.


\section{Conclusions}

We have introduced a response-space formulation of the
finite-basis exchange-only optimized effective potential in
which physically motivated response functions are used as
adaptive auxiliary directions. The BJ, RPP, GLLB, KLI, and LHF
constructions are retained as sources of spatial information,
while their amplitudes or occupied-space coefficients are
determined directly from the projected OEP equation. This
provides a systematic connection between conventional model
exchange potentials and finite-basis OEPx calculations.

Adapting the amplitude of a single BJ or GLLB response direction
improves both the exchange-only total energies and the
HOMO-based ionization potentials, although the accuracy is
ultimately limited by the one-dimensional form of these spaces.
Adding the independent RPP direction in aRG provides additional
flexibility and further reduces both errors. A more substantial
improvement is obtained with the occupied-orbital aKLI and
occupied-pair aLHF expansions. The aKLI construction gives a
compact and generally well-conditioned representation, while
aLHF provides the closest agreement with the finite-basis OEPx
reference among the spaces considered here. 

The comparison of the self-consistent potentials and densities
for Ne and CO confirms that this improvement is not restricted
to integrated energetic quantities. The adaptive response spaces
recover the dominant shell, intershell, and bonding structures
of the OEPx response with far fewer directions than a generic
auxiliary expansion. Within the spatial ranges represented by
the finite orbital bases, the most flexible adaptive potentials
also closely reproduce the gauge-aligned OEPx tails and the
associated density response.

The response spectra reveal, however, that increasing accuracy
and increasing numerical flexibility are accompanied by a
conditioning cost. The occupied-pair dimension of aLHF grows
quadratically with the number of occupied orbitals and can
generate near-null response modes, as observed for CO. Stable
solutions therefore require an explicit TSVD treatment and
monitoring of both the retained rank and the retained-space
condition number. By contrast, aRG and aKLI remain substantially
smaller and better conditioned. The results consequently suggest
that the most useful response space is not necessarily the
largest one, but the smallest physically adapted space that
captures the relevant structure of the OEP response without
introducing poorly identifiable directions.

Several natural extensions of the present work can be envisaged. The response-space strategy may be generalized to orbital-dependent correlation functionals, for which the construction of stable and computationally efficient local potentials remains a significant challenge~\cite{LHFc}. Another promising direction is the application of aBJ and related adaptive response-space methods to periodic systems and solid-state calculations. These developments will be pursued in future work.

\section{Dedication}

None of us had the privilege of meeting Alex Becke personally, yet his ideas have been a quiet and enduring presence throughout our scientific journey. His work transformed the way exchange is understood and constructed within density-functional theory, demonstrating how profound physical insight can render seemingly formidable problems into elegant and useful approximations.

The BJ construction exemplifies this philosophy: a concise expression, rooted in fundamental principles, that captures intricate features of the exact-exchange potential. The present work, which seeks to understand and adapt the response component of that potential, is deeply inspired by this same spirit.

We dedicate this work to the memory of Alex Becke, with gratitude for the ideas he gave to our field, admiration for the elegance of his scientific contributions, and respect for the lasting influence of his work on generations of researchers.

\section*{Acknowledgements}

S.Ś. acknowledges partial
financial support from the National Science Centre,
Poland, under Grant No. 2025/59/B/ST4/00523.

\section*{Data Availability}

All data supporting the findings are deposited in the Zenodo repository~\cite{adaptive2026}. Additional data are available from the corresponding author upon reasonable request.


\appendix
\twocolumngrid

\bibliographystyle{apsrev4-1}
\bibliography{biblio_fixed}

\end{document}